\documentclass{aastex}

\newcommand{\kms}{km~s$^{-1}$}
\newcommand{\ie}{{\it i.e., }}
\newcommand{\etal}{{\it et al. }}
\newcommand{\eg}{{\it e.g., }}

\shorttitle{Radio Emission from SN~1994I}
\shortauthors{Weiler, \etal}

\begin{document}

\title{Radio Emission from SN~1994I in NGC 5194 (M 51) - The Best Studied Type Ib/c Radio Supernova} 

\author{Kurt W. Weiler}
\affil{Naval Research Laboratory, Code 7213, Washington, DC 20375-5320; Kurt.Weiler@nrl.navy.mil}

\author{Nino Panagia\altaffilmark{1}}
\affil{Space Telescope Science Institute, 3700 San Martin Drive, Baltimore, MD 21218; panagia@stsci.edu}


\author{Christopher Stockdale}
\affil{Physics Department, Marquette University, P.O. Box 1881, Milwaukee, WI 53201-1881; Christopher.Stockdale@marquette.edu}

\author{Michael Rupen}
\affil{National Radio Astronomy Observatory, P.O.~Box 0, Socorro, NM 87801; mrupen@nrao.edu}

\author{Richard A. Sramek}
\affil{National Radio Astronomy Observatory, P.O.~Box 0, Socorro, NM 87801; dsramek@nrao.edu}

\and

\author{Christopher L.~Williams}
\affil{MIT Kavli Institute for Astrophysics and Space Research, Cambridge, MA 02139; clmw@mit.edu}

\altaffiltext{1}{INAF-CT Osservatorio Astrofisico di Catania, Via S. Sofia 79, I-95123 Catania, Italy; Supernova Ltd, OYV \#131, Northsound Rd, Virgin Gorda, British Virgin Islands}

\begin{abstract}

We present the results of detailed monitoring of the radio emission from the Type Ic supernova SN~1994I from 3 days after optical
discovery on 1994 March 31 until eight years later at age 2927 days on 2002 April 05.  The data were mainly obtained using the
Very Large Array at the five wavelengths of $\lambda \lambda$1.3, 2.0, 3.6, 6.2, and 21 cm and from the Cambridge
5 km Ryle Telescope at $\lambda$2.0 cm.  Two additional measurements were obtained at millimeter wavelengths. This data set represents the most complete, multifrequency radio observations ever obtained for a Type Ib/c supernova. The radio emission evolves regularly in both time and frequency and is well described by established SN emission/absorption models. It is the first radio supernova with sufficient data to show that it is clearly dominated by the effects of synchrotron self-absorption at early times.
\end{abstract}


\section{Introduction\label{intro}}

\subsection{Background\label{background}}

By 1941 \cite{Minkowski41} had recognized at least two classes of SNe -- Type I with no hydrogen in their optical spectra and Type II with hydrogen lines in their optical spectra. By 1980 radio observations of supernovae began to play a role with the clear detection of SN~1979C \citep{Weiler80} (there had been only one earlier detected radio supernova (RSN), SN~1970G, by  \citealt{Gottesman72,Goss73} but the Westerbork telescope was unable to cleanly separate its radio emission from a nearby HII region) and the monitoring of its radio light curves at multiple radio frequencies for many years thereafter \citep{Montes00}. The radio detection and monitoring of SN~1980K quickly followed \citep{Weiler86,Montes98}. However, both objects were optically identified as Type II SNe. At that time, little was known about the radio emission from SNe and the statistics with only two objects were very poor. However, the radio detection of SN~1983N \citep[][also labelled SN~1983.51 by those authors]{Sramek84}, SN~1984L \citep{Panagia86}, and SN~1990B \citep{VanDyk93} appeared to establish radio detections of Type I SNe. Around that time, \cite{Panagia86} proposed that, from their optical spectra, Type I SNe are not a homogeneous class but split into at least two sub-classes. The appearance of SN~1994I, the subject of this paper, finally yielded a bright, nearby object for detailed study of a ``radio loud'' Type I supernova.  

Currently, our understanding of the classes of SNe is such that the category of Type I supernovae is clearly bifurcated. One now speaks of Type Ia SNe  \citep[which still have never been detected in the radio, see][]{Panagia06} as originating in the deflagration of small stars, such as accreting or merging white dwarfs or black holes, and Type Ib/c and Type II (of which there are also numerous subclasses) SNe as originating in the core collapse of massive stars with a ZAMS masses of $>8$ M$_\odot$.

Particularly since the discovery of an association of SN~1998bw with GRB~980425 \citep{Sadler98}, the study of Type Ib/c SNe has greatly increased (as explained  below, the optical spectroscopic differences between Type Ib and Type Ic SNe are so slight that we treat them as a single class). However, further work has found that most Type Ib/c SNe ($\ge$97\%) are not associated with detectable GRBs \citep{Berger03a,Soderberg06}.

Even with the enhanced study of Type Ib/c SNe since SN~1998bw, most are still relatively poorly sampled in the radio \citep[see, e.g.,][]{Berger03b,Soderberg04,Soderberg05,Soderberg10} and, even at the present time, no Type Ib/c SN has been as thoroughly studied in the radio as SN~1994I presented in this paper.

\subsection{SN~1994I\label{sn1994I}}

SN~1994I in M~51 (NGC~5194) was discovered at magnitude $V = 13{\fm}5$ on 1994 April 2.17 UT \citep{Puckett94}
with independent discoveries by other observers following quickly (see IAUC 5961).  \citet{Richmond94a}
confirmed the discovery as a supernova (SN) and noted it as being ``very blue.''  \citet{Armus94}
obtained moderate-dispersion spectra and noted ``broad undulations superimposed on a generally featureless continuum.''

Initial confusion arose as to the type identification of the SN with \citet{Schmidt94a} first (incorrectly)
identifying it as a Type II SN based on unreduced spectra taken by J. Peters and G. Bernstein during
 1994 April 3.37-3.52 UT.  The spectra, however, did suggest the SN location in NGC~5194 (M51) by
identifying strong Na D lines at the recession velocity of the galaxy. By 1994 April 4.4 UT, \citet{Filippenko94}
suggested from new spectra that SN~1994I was a Type Ib SN.  However, they noted the ``possible presence of
weak H$_\alpha$'' and suggested that, although the presence of hydrogen would ``technically make it a Type
IIb,'' the steep drop of the continuum toward the blue and near-ultraviolet was more consistent with a Type Ib
identification, as were other aspects of the spectrum.  Finally, \citet{Schmidt94b} confirmed that SN~1994I was
not a Type II SN and \citet{PhillipsM94} pointed out the spectral similarities to the Type Ic SN~1983V.  The Type
Ic identification was also confirmed by \citet{Clocchiatti94}.  (Note: Because the spectral differences between the SN Type Ib and Type Ic
optical classes are slight -- Type Ib show strong He I absorption while Type Ic show weaker He I absorption  -- and
there are no obvious radio differences, we shall hereafter refer to Type Ib/c SNe as one class.)  \citet{Richmond94b}
discovered a red star near the SN position on HST PC images from 1992 which he suggested as a possible progenitor to
the SN, although his suggestion is disputed by \citet{Kirshner94}.

Already on 1994 April 3.17 UT, \citet{Rupen94} detected radio emission at the position of SN~1994I with
the VLA\footnote{The VLA telescope of the National Radio Astronomy Observatory is operated by Associated
Universities, Inc. under a cooperative agreement with the National Science Foundation.}
at both $\lambda$1.3 cm (22.4 GHz) and $\lambda$3.6 cm (8.4 GHz).  The radio observations
established an accurate position of RA(J2000) = $13^h 29^m 54\fs12$, Dec(J2000) = $+47\arcdeg 11\arcmin 30\farcs4$
with an uncertainty of $\pm$ 0\farcs1 in each coordinate with an offset from the radio nucleus of M 51 of $14\farcs3$
east and $12\farcs3$ south.  \citet{Morrison94} report a slightly different optical
astrometric position of RA(J2000) = $13^h 29^m 54\fs072$, Dec(J2000) = $+47\arcdeg 11\arcmin 30\farcs50$
with uncertainty of $\pm$ 0\farcs08 in each coordinate, a difference of $\sim0\farcs5$ in RA and $\sim0\farcs1$
in Dec.  This optical/radio position difference might imply that there is a significant offset between the
centroids of the optical and radio emission (unlikely since, at the distance of M51 at 8.9 Mpc, $0\farcs5$ is $21.5$ parsecs)
or, more likely, that there is an offset between the radio reference frame and the FK5 based optical reference
frame in that part of the sky.
\citet{Tully88} determined a distance of 7.7 Mpc (assuming H$_{\rm 0}$ = 75 km s$^{-1}$ Mpc$^{-1}$)
to NGC~5194. Accepting the \citet{Tully88} method and using 
H$_{\rm 0}$ = 65 km s$^{-1}$ Mpc$^{-1}$, we will adopt 8.9 Mpc as the distance to SN~1994I.  Other distances to
M51 range from 9.6 Mpc (H$_{\rm 0}$ = 57; derived from HII regions and brightest stars by \citealt{Sandage74}) to 8.4 Mpc \citep{Feldmeier97} from planetary nebula luminosity.

Ultraviolet measurements were reported from IUE on 1994 April 3.35, 4.9, and 5.13 by \citet{Rodriguez94} and an infrared detection was reported on 1994 April 23.2 UT by \citet{Rudy94} in the J, H, and 2.00-2.32 $\micron$ windows. An unabsorbed soft X-ray luminosity upper limit of $<1.5~\times$ 10$^{38}$ erg s$^{-1}$ for a distance to M51 of 7.7 Mpc was reported by \citet{Lewin94} from the ROSAT satellite for 1994 May 22 and subsequent X-ray studies yielded
the detection of SN~1994I with ROSAT at 82 days after explosion with $(1.82\pm 4.2) \times \ 10^{38}$ erg s$^{-1}$ \citep{Immler98}. The CHANDRA X-ray Facility at 2271 and 2639 days  after explosion reported $(1.4\pm 0.7) \times \ 10^{37}$ erg s$^{-1}$ and $(1.2\pm 0.5) \times \ 10^{37}$ erg s$^{-1}$, respectively \citep{Immler02}.  \citet{Immler02} determined a best-fit X-ray rate of decline proportional to $t^{-1}$ and a circumstellar medium
(CSM) density profile of $\rho_{\rm CSM} \propto r^{-1.9\pm 0.1}$, consistent a constant mass-loss rate/constant wind velocity for the SN progenitor which would yield a CSM density profile of $\rho_{\rm CSM} \propto r^{-2}$.

Prediscovery limits are also available. \citet{Richmond94c} reports that SN~1994I was not present to a limiting R
magnitude of $\sim$16.2 on 1994 March 28 UT and a prediscovery measurement with the VLA by \citet{Tongue94} found
a 3$\sigma$ upper limit for $\lambda$21 cm emission from that position of $<$2.3 mJy on 1994 February 12.

Due to its proximity (8.9 Mpc; Tully 1988 recalculated for H$_{\rm 0}$ = 65 km s$^{-1}$ Mpc$^{-1}$) and the fact that such an early radio detection was obtained, we quickly started observations at all available wavelengths with the VLA. Regular monitoring of SN~1994I with
the VLA at $\lambda \lambda$1.3 cm (22.5 GHz), 2 cm (14.9 GHz), 3.6 cm (8.4 GHz), 6 cm (4.9 GHz), and 20 cm
(1.4 GHz) then continued until the end of 1994 with sporadic measurements thereafter. \citet{Pooley94} also monitored the SN extensively at 2 cm (15.2 GHz) with the Ryle Telescope. These data are listed in Table \ref{tbl-1}. Two additional observations were taken with the Caltech OVRO millimeter interferometer at 3 mm (99 GHz) and at 1.4 mm (218 GHz)  \citep{PhillipsJ94}. These are listed in Table \ref{tbl-2} .  

\section{Radio Observations\label{radioobs}}

\subsection {Flux density measurements\label{fluxdensities}}

The VLA radio observations reported here were made between 1994 April 03 and 2002 April 05. VLA phase and flux density calibration and data reduction followed standard procedures \citep[see, e.g., ][]{Weiler86, Weiler90}, using 3C~286 as the primary flux density calibrator and J1418$+$546 as the secondary flux density and phase calibrator. As noted in the footnotes to Table \ref{tbl-3}, two measurements on 1994 April 03 used the calibrator J1415$+$463 and one measurement on 2002 April 05 used the calibrator J1327$+$434. Table \ref{tbl-3} lists the measured or interpolated (or extrapolated) flux densities of the secondary calibrators. For the phase calibration,  J1327$+$434, J1415$+$463, and J1418$+$546 have defined positions as listed in the NRAO VLA calibrator catalogue.

The flux density measurements for SN~1994I are listed in Tables \ref{tbl-1} and \ref{tbl-2}  and plotted in Figure \ref{fig-1} with the filled squares
representing actual detections and inverted open triangles representing 3$\sigma$ upper limits. 
Also shown in Figure \ref{fig-1} are the best fit model curves discussed below in Section \ref{radiomodels} and Section \ref{discussion} with
the set of parameters reported in Table 4.

The flux density measurement error given for the VLA measurements in Table \ref{tbl-1} is a combination of the
rms map error, which measures the contribution of small unresolved fluctuations in the background emission
and random map fluctuations due to receiver noise, and a basic fractional error $\epsilon$
included to account for the normal inaccuracy of VLA flux density
calibration and possible deviations of the primary calibrator from an
absolute flux density scale.  These final errors ($\sigma_f$) are taken as

\begin{equation}
\label{eq1}
\sigma_{f}^{2} = (\epsilon S_0)^2+\sigma_{0}^2 
\end{equation}

\noindent where $S_0$ is the measured flux density, $\sigma_0$ is the map rms for each
observation, and $\epsilon =0.10$ for 20 cm, $\epsilon =0.05$ for 6 and 3.6 cm,  $\epsilon
= 0.075$ for 2 cm, and $\epsilon =0.10$ for 1.2 cm. For all of the Cambridge measurements at 2 cm, $\epsilon =0.10$ was assumed.

\subsection {Spectral indices\label{spectralindices}}

With such extensive data, we can also calculate the spectral index evolution for SN~1994I and these are shown in Figure  \ref{fig-2} for four pairs of
bands (1.2-2\,cm, 2-3.6\,cm, 3.6-6\,cm, and 6-20\,cm). The parameters listed in Table 4 from the best fit light curves allow calculation of the model spectral index evolution curves and these are also shown in Figure \ref{fig-2}.  At the earliest times, the observed
spectral indices exceed the maximum value spectral index ($\alpha = 2.5$) possible for a purely synchrotron self-absorption source,  suggesting that at least some thermal absorption is present in SN~1994I.

The overall model fits to the spectral index evolution are quite good, with the only exception being the model
1.2-2\,cm spectral index at early times, which is systematically higher than
the observed values.  This deviation, which was found also for the case of
SN\,1993J \citep{Weiler07}, suggests that, at the highest frequencies, the intrinsic radio spectrum
has a somewhat flatter spectral index than the best fit value determined by the longer wavelength data.

\subsection{Brightness temperature\label{brightnesstemp}}

This extensive data set can also be used to estimate the intrinsic brightness temperature of the SN~1994I radio emission which can be derived
from the ratio of the observed radio flux density to the solid angle $\Omega$ of the emitting region ($T_B\propto S_{obs} /
\Omega$). 

Since SN\,1994I was not radio bright enough, it was not possible to directly measure its
angular size by means of VLBI observations, as had been possible for SN~1993J \citep[see, e.g.,][and references therein]{Marcaide09}, but the physical size of the SN shock can be estimated from the optically measured expansion velocity and distance to the SN.  For SN~1994I, published expansion velocities
at early times range between $\sim$16,500 \kms ~\citep{Filippenko95,Millard99} from the absorption troughs of He I $\lambda ~10830$, through $\sim$20,000 \kms
~\citep{Millard99} as seen for Si II, up to almost 30,000 \kms ~that is the
extent measured by \cite{Filippenko95} for the blue wing of the He I $\lambda ~10830$ absorption line.  We shall adopt a compromise expansion velocity of 20,000 \kms and, following \cite{Immler02}, we shall assume that it is constant with time.  Therefore, the
radius $\rm r$ of the emitting region can be estimated as:

\begin{equation}
\label{eq2}
r = 1.73\times10^{14} ~(\rm t/1 ~\rm day) ~~\rm cm
\end{equation} 

\noindent which,
for an adopted distance D to SN~1994I of
8.9\,Mpc, gives an angular size of 

\begin{equation}
\label{eq3}
\theta = 1.29 ~(\rm t/1 ~\rm day)~~\mu \rm arcsec
\end{equation} 

\noindent and an estimated brightness temperature of 

\begin{equation}
\label{eq4}
T_B = 3.5\times 10^{11} S_{\rm obs}(\rm mJy)~ \lambda ^2 ~(\rm t/1 ~\rm day)^{-2} ~\rm K
\end{equation}

The resulting brightness temperature evolution is shown in Figure \ref{fig-3} (left). The corresponding brightness temperature evolution curves for SN~1993J from \cite{Weiler07} are also shown in Figure \ref{fig-3} (right).
 
From these brightness temperature estimates, we can constrain the value of external absorption due to a presupernova stellar wind by requiring that $T_B$ not exceed the limit of 
$\sim 3\times10^{11}$~K \citep{Kellermann69,Readhead94} where the radio emission is quenched by inverse-Compton scattering. This implies that any value of $K_2$ (see Section \ref{parameterization}) higher than $\sim30$ would
yield $T_B$ curves in strong disagreement with such a physical requirement and puts an upper limit on the amount of external thermal absorption by a presupernova mass-loss in a dense stellar wind.

\section{Radio supernova models\label{radiomodels}}

\subsection{Parameterized model\label{parameterization}}

\citet{Weiler86} discuss the common properties of radio SNe (RSNe), including non-thermal
synchrotron emission with high brightness temperature, turn-on delay at longer wavelengths,
power-law decline with index $\beta$ after maximum, and spectral index $\alpha$ (S $\propto \nu^{+\alpha}$) asymptotically
decreasing to an optically thin value.  \citet{Weiler86,Weiler90} have shown that the ``mini-shell'' model
of \cite{Chevalier82a,Chevalier82b}, with modifications by \citet{Weiler90}, adequately describes
previously known RSNe.  In this model, the relativistic electrons and enhanced magnetic fields necessary
for synchrotron emission are generated by the SN shock interacting with a relatively high-density circumstellar
envelope which has been ionized and heated by the initial UV/X-ray flash.  Figure \ref{fig-4} (taken from \cite{Weiler02}) shows a, not to scale, greatly simplified cartoon of the vicinity of a supernova explosion. This dense cocoon is presumed to have
been established by a constant mass-loss ($\dot M$) rate, constant velocity ($w$) wind (i.~e., $\rho_{\rm CSM} \propto r^{-2}$)
from a massive ( $>8$ M$_\odot$) stellar progenitor or companion.  This ionized CSM can also be the source of initial thermal absorption (Free-Free Absorption, FFA) although \citet{Chevalier98} has suggested that non-thermal, synchrotron self-absorption (SSA) may be a significant source of absorption in the early phases of some RSNe. For the case of FFA, the
rapid rise in radio flux density results from the shock overtaking progressively more of the wind matter, leaving less of it
along the line-of-sight to the observer to absorb the more slowly decreasing synchrotron emission from the shock region. In the case of SSA, the non-thermal absorption decreases as the emitting region expands and decreases in density. \citet{Montes97} also allow for
 the possibility of intervening HII along the line-of-sight to the RSN, providing a constant thermal absorption of
  the radio emission.

To parameterize the basic characteristics of radio supernovae, we adopt the most recent RSN discussion of \citet{Weiler02}
and \citet{Sramek03}\footnote{Even though this model has been previously described in papers such as the references given above, we list it again for completeness, for the reader's convenience, and for explanation of the parameters listed in Table~\ref{tbl-4}} and illustrate the regions of the supernova being parameterized in a cartoon shown in Figure \ref{fig-4} from, e.g., \citet{Weiler02} :

\begin{equation}
\label{eq5}
S(\mbox{mJy}) = K_1 \left(\frac{\nu}{\mbox{5\
GHz}}\right)^{\alpha} \left(\frac{t-t_0}{\mbox{1\ day}}\right)^{\beta}
e^{-\tau_{\rm external}} \left(\frac{1-e^{-\tau_{{\rm CSM}_{\rm clumps}}}}{\tau_{{\rm CSM}_{\rm
clumps}}}\right) \left(\frac{1-e^{-\tau_{\rm internal}}}{\tau_{\rm internal}}\right) 
\end{equation} 

\noindent with  

\begin{equation}
\label{eq6}
\tau_{\rm external}  =  \tau_{{\rm CSM}_{\rm homogeneous}}+\tau_{\rm distant},
\end{equation}

\noindent where

\begin{equation}
\label{eq7}
\tau_{{\rm CSM}_{\rm homogeneous}}  =  K_2
\left(\frac{\nu}{\mbox{5 GHz}}\right)^{-2.1}
\left(\frac{t-t_0}{\mbox{1\ day}}\right)^{\delta}
\end{equation}

\begin{equation}
\label{eq8}
\tau_{\rm distant}  =  K_4  \left(\frac{\nu}{\mbox{5\
GHz}}\right)^{-2.1}
\end{equation} 

\noindent and

\begin{equation}
\label{eq9}
\tau_{{\rm CSM}_{\rm clumps}}  =  K_3 \left(\frac{\nu}{\mbox{5\
GHz}}\right)^{-2.1} \left(\frac{t-t_0}{\mbox{1\
day}}\right)^{\delta^{\prime}}
\end{equation} 

\noindent with $K_1$, $K_2$, $K_3$, and $K_4$ determined from fits to the data and corresponding, formally, to the flux density ($K_1$), homogeneous ($K_2$, $K_4$), and clumpy or filamentary ($K_3$) FFA 
at 5~GHz one day after the explosion date $t_0$.  The terms $\tau_{{\rm CSM}_{\rm homogeneous}}$ and  $\tau_{{\rm CSM}_{\rm clumps}}$ describe the attenuation due to local, homogeneous FFA CSM and clumpy or filamentary FFA CSM, respectively, that are near enough to the SN progenitor that they are altered by the rapidly expanding SN blastwave.  The $\tau_{{\rm CSM}_{\rm homogeneous}}$ FFA is produced by an ionized medium that completely covers the emitting source (``homogeneous external absorption''), and the $(1-e^{-\tau_{{\rm CSM}_{\rm clumps}}}) \tau_{{\rm CSM}_{\rm clumps}}^{-1}$ term describes the attenuation produced by an inhomogeneous FFA medium \citep[``clumpy absorption''; see][for a more detailed discussion of attenuation in inhomogeneous media]{Natta84}. The $\tau_{\rm distant}$ term describes the attenuation produced by a homogeneous FFA medium which completely covers the source but is so far from the SN progenitor that it is not affected by the expanding SN blastwave and is consequently constant in time.  All external and clumpy absorbing media are assumed to be purely thermal, singly ionized gas which absorbs via FFA with frequency dependence $\nu^{-2.1}$ in the radio.  The parameters $\delta$
and $\delta'$ describe the time dependence of the optical depths for the local homogeneous and clumpy or filamentary media, respectively. 

Since it is physically realistic and is needed in some RSNe where radio observations have been obtained at early times and high frequencies, Equation \ref{eq5} also includes the possibility for an internal absorption term\footnote{Note that for simplicity an internal absorber attenuation of the form $\left(\frac{1-e^{-\tau_{{\rm CSM}_{\rm internal}}}}{\tau_{{\rm CSM}_{\rm internal}}}\right)$, which is appropriate for a plane-parallel geometry, is used instead of the more complicated expression \citep[e.g., ][]{Osterbrock74} valid for the spherical case.  The assumption does not affect the quality of the analysis because, to within 5\% accuracy, the optical depth obtained with the spherical case formula is simply three-fourths of that obtained with the plane-parallel slab formula.}.  This internal absorption  ($\tau_{\rm internal}$) term may consist of two parts -- synchrotron self-absorption (SSA; $\tau_{{\rm internal}_{\rm SSA}}$), and thermal FFA ($\tau_{{\rm internal}_{\rm FFA}}$) due to ionized gas mixed with non-thermal emission. 

\begin{equation}
\label{eq10}
\tau_{\rm internal}  = \tau_{\rm internal_{\rm SSA}} + \tau_{\rm internal_{\rm FFA}}
\end{equation}

\begin{equation}
\label{eq11}
\tau_{\rm internal_{\rm SSA}} = K_5\left(\frac{\nu}{\mbox{5\
GHz}}\right)^{\alpha-2.5}  \left(\frac{t-t_0}{\mbox{1\
day}}\right)^{\delta^{\prime\prime}}
\end{equation}

\begin{equation}
\label{eq12}
\tau_{\rm internal_{\rm FFA}}  =   K_6  \left(\frac{\nu}{\mbox{5\
GHz}}\right)^{-2.1} \left(\frac{t-t_0}{\mbox{1\
day}}\right)^{\delta^{\prime\prime\prime}}
\end{equation}

\noindent with $K_5$ corresponding, formally, to the internal,
non-thermal ($\nu^{\alpha - 2.5}$) SSA and $K_6$, corresponding, formally,
to the internal thermal ($\nu^{-2.1}$) FFA mixed with
non-thermal emission, at 5~GHz one day after the explosion date $t_0$. 
The parameters $\delta^{\prime \prime}$ and $\delta^{\prime \prime
\prime}$ describe the time dependence of the optical depths for the SSA
and FFA internal absorption components, respectively. 

Application of this basic parameterization has been shown to be effective
in describing the physical characteristics of the presupernova system,
its CSM, and its final stages of evolution before explosion for objects
ranging from the two decades of monitoring the complex radio emission
from SN~1979C \citep{Montes00}, through the unusual SN~1998bw (GRB980425)
\citep{Weiler01b}, to most recent $\gamma$-ray bursters/Type Ib/c SNe 
\citep{Weiler02,Weiler03}.


\section{Discussion\label{discussion}}

\subsection {Application of models to SN~1994I\label{apply1994I}}

As was clearly demonstrated by Weiler et al (2007) for the case of SN\,1993J, in
order to obtain a meaningful representation of radio observations of a supernova, a
model must be able to account simultaneously for three different aspects,
namely: (a) the light curves measured at several epochs and in several distinct
radio bands, (b) the corresponding spectral indices computed for pairs of bands at
adjacent frequencies, and (c) the brightness temperatures, which at all times
should be lower than the well established limit of $3\times 10^{11}$K
(Kellermann \& Pauliny-Toth 1969, Readhead 1994).

\subsection {SN~1994I radio light curves\label{lightcurves}}

We have excellent time coverage of the radio emission from SN~1994I at 5 frequencies, from 22~GHz down
to 1.4~GHz.  That enables us to make a detailed study of its radio properties. In particular, the rising branch of the light curve at early times,
where absorption ($\tau>>1$) is still  important, is well defined in all five
main bands. Also, the flux density evolution is well characterized by a power law decay at late
times, in the optically thin phases.  Therefore, we are in an excellent
position to identify the main processes operating in this source and determine
their parameters.  Actually, the best fit curves as described by the parameters in Table~\ref{tbl-4} and shown in Figure~\ref{fig-1} agree quite well with the
observations, with an overall $\chi^2 = 4.42$ per degree of freedom. 

From this rather good parameterization of the radio emission from SN~1994I, we
conclude that the radio light curves are not well described by a purely thermal
absorption model such as that given by Equations \ref{eq5} to \ref{eq9}.  In particular, the
requirement for  non-zero $K_5$ and $K_6$ parameter values implies the existence
of significant synchrotron self-absorption (SSA). In fact, the early optical
depth is dominated almost entirely by the SSA ($K_5$) component and the FFA component ($K_6$)
makes a significant contribution only as the emission becomes more optically
thin and the brightness temperature begins to drop. As predicted by  \cite{Chevalier98} there does not appear to be
a highly-structured, thick-CSM  present since there is no need for
a large $K_2$ or non-zero $K_3$ term. The necessity of a $K_4$ term 
(\ie thermal absorption by foreground HII gas; see Section \ref{externalabsorption}) is consistent with
recent deep 20~cm observations that indicate diffuse radio emission from the
host galaxy in the region near the supernova \citep{Maddox07}. Further
evidence for substantial amounts of HII gas  is provided by the 
H$\alpha$ emission detected on HST images of the immediate vicinity of  SN~1994I
\citep{Maddox06,Maddox07}.   

It is interesting to note that SN~1994I shows a dip around day 27 in its
flux density at centimeter wavelengths (frequencies from 4.9 GHz to 22
GHz).  While less pronounced than that noted for the unusual Type Ib/c
SN~1998bw, which is related to the $\gamma$-ray burster GRB 980425, it occurs on
a similar time scale \citep[see][and {\it
http://www.narrabri.atnf.csiro.au/public/grb/grb980425/}]{Kulkarni98}.  This
early ``dip'' followed by ``re-brightening'' for SN~1998bw has variously been
explained as due to a slowing and re-energization of the SN blast wave
\citep{Li99} or, more likely, as structure in the  circumstellar absorbing
material \citep{Weiler01a, Weiler01b}.

Considering the light curves shown in Figure \ref{fig-1} in more detail, even though the fitted model provides a good representation the agreement is never quite perfect, mostly because the model is rather
simplified and also because there may be systematic deviations
from our basic assumptions.  In order to analyze the residuals
between the model and the observations, we express them as logarithms of the
ratios of observed to model fluxes and plot them in Figure~\ref{fig-5} (left).  A similar analysis helped us in other
RSNe to reveal unsuspected features such as a periodic flux modulation for
SN~1979C \citep{Weiler92,Schwarz96} possibly due to the presence of a companion,
as well as evidence for interacting winds in SN\,1998bw \citep{Weiler01b} and 
SN~2001ig  \citep{Ryder04}.

The residuals for SN\,1994I are displayed separately for individual bands in Figure~\ref{fig-5} (left), and merged in a single plot in Figure \ref{fig-5} (right).  As expected, they show scatter inherent to non-negligible measurement errors and to a simplified
description of a complicated phenomenon. However,  a systematic feature is
apparent, essentially with the same timing and
comparable amplitudes, at all bands.  In particular, in Figure \ref{fig-5} (right) we can
see a somewhat broad trough, with an apparent local minimum, which is about 0.2 deep in the
logarithm and occurring $\sim 27$ days after explosion. This trough extends from
$\sim 10$ up  to $\sim 55$ days after explosion.  Since this local minimum is present at
all frequencies with similar amplitude, it must represent a genuine decrease in
the intrinsic emission rather than being the result of additional, temporary 
absorption along the line of sight.  In a spherically symmetric model it
would correspond to a temporary decrease of the average density of the CSM by
$\sim 0.1$ in the logarithm of the ratio,  or a factor of $\sim 0.8$ decrease in the CSM density, with a duration of
$\sim 45$ days.  

Although a temporary decrease in the mass-loss rate from the presupernova star cannot be ruled out, we feel it worthwhile to also consider a model in
which the SN~1994I progenitor was a Wolf-Rayet (WR) star in a wide binary system whose companion was
also an early type star (most likely a  B-type star on the main sequence). Such a binary system could create  non-spherical CSM density structures by the interacting winds of the system
stars \citep[see, e.g.,][]{Schwarz96}. If such were the case, for SN~1994I the observed  minimum in the residuals would correspond to a separation
between the two stars of $\sim4.7\times10^{15}$ ~cm or $\sim310$~AU [see also SN~1979C \citep{Weiler92,Schwarz96} where the radio emission modulation is possibly due to the presence of a companion, as well as evidence for interacting winds in SN\,1998bw \citep{Weiler01b} and 
SN~2001ig  \citep{Ryder04}].

\subsection {SN~1994I spectral index evolution \label{spectralindexevolution}}

Examination of the spectral index evolution in Figure \ref{fig-2} shows three things: 1) The model listed in Table \ref{tbl-4} generally describes the spectral index evolution quite well except at at 1.2-2 cm. Apparently, at higher frequencies the model is beginning to fail, 2) The spectral index evolution, particularly at the first part at 3.6-6 cm,  requires some (small) amount of thermal absorption since synchrotron self-absorption (SSA) can never generate a spectral index exceeding +2.5 ; and 3) There is no evidence for systematic spectral index change of the emission within the available observations. While short term fluctuations of the spectral index cannot be ruled out, none appear to occur at the same epoch at all frequencies.

\subsection {Brightness temperature\label{brightnesstemperature}}

Figure~\ref{fig-3} shows the estimated brightness temperatures for each
frequency versus time for SNe~1994I (left) and for 1993J (right; 
from \citealt[][]{Weiler07}).  While the precise temperature measurements depend on
the distance used and the assumed (or, for SN~1993J, measured) shock wave speed,
it is clear that the two plots have a systematic difference at early times.  In
the case of SN~1994I, the temperatures for each frequency appear to remain
approximately constant near the maximum possible value of $3 \times 10^{11}$ K (see Section \ref{brightnesstemp} until the source becomes optically thin.  However, the
measured brightness temperatures for SN~1993J at the earliest times  increase
for each frequency until the emission becomes optically thin. This is 
evidence that the dominant absorption  mechanism is different between these two SNe,
with the primary absorption
mechanism for SN~1993J not directly related to the emission mechanism (\ie
external thermal electrons), while the primary absorption mechanism  for SN~1994I is physically associated with
the emission mechanism (\ie synchrotron electrons).  This simple comparison makes it clear that SSA is
dominant in SN~1994I at early times while, as demonstrated by \cite{Weiler07},
SN~1993J requires significant contributions to the early absorption from {\it both}
FFA and SSA.  Thus, SN~1994I is the first observed radio supernova (RSN) to clearly
demonstrate the need for a SSA dominated component to model its radio light
curves.

\subsection {External distant absorption\label{externalabsorption}}

The parameter $K_4$ (see Section \ref{parameterization}) is interpreted as being due to a constant amount of
ionized material between us and SN~1994I that is sufficiently far from the SN to
be unaffected by the expanding shock wave. It is related to the emission measure (EM) of this intervening ionized hydrogen by Equation (1-223) of
\cite{Lang86} so that

\begin{equation}
\label{eq13}
{\rm EM} = 8.93 \times 10^7 ~{\rm K_4}~({\rm T_e}/10^4
{\rm K})^{1.35}  {\rm pc}~ {\rm cm}^{-6} 
\end{equation} 

\noindent where ${\rm T_e}$ is the electron
temperature of the ionized absorbing medium.  Thus, the value for ${\rm K_4}$ in
Table~\ref{tbl-4} implies a presence along the line of sight to SN~1994I of ionized
hydrogen with an emission measure of ${\rm EM} \simeq 4.1\times 10^6~({\rm
T_e}/10^4~ {\rm K})^{1.35}$~pc cm$^{-6}$. 

Similar to the case of SN~1978K \citep{Montes97,Chu99} this  thermal absorption 
is likely to be due to an intervening H II region and not part of the matter
from the presupernova mass-loss wind. Most HII regions with an emission measure
higher than $\sim 10^6$~pc~cm$^{-6}$ are known to be rather  compact sources
\citep[see, \eg][]{Wilson70,Reifenstein70}; for example, the Orion Nebula has 
an average emission measure of $1.8 \times 10^6$~pc~cm$^{-6}$ and a diameter of
$\sim 2$~pc \citep{Allen73}.  If we adopt a similar diameter of $\sim 2$ pc for this HII region presumed to be along the line-of-sight to SN~1994I, we derive an average electron density of $2\times 10^3$~pc~cm$^{-6}$ and a
total mass of the ionized gas of ${\rm M(HII)} \simeq 250~ {\rm M}_\odot$. The  volume
emission measure is then $n_e^2V\simeq 5.1\times 10^{62}$~cm$^{-3}$,
which implies a steady source of ionizing radiation of $N_L\simeq 1.3\times
10^{50}$ Lyman-continuum photons~s$^{-1}$. Such a flux is higher than that
provided even by a bright O-type star \citep[see, \eg][]{Panagia73} and suggests
the presence of a young, compact star cluster in the foreground of SN~1994I.  

In order not to make the projection of such a powerful cluster in front of
SN~1994I too geometrically improbable, the cluster should be spatially close to the supernova
itself; say, separated by no more than $\sim 3$~pc,  which would give a
probability of chance projection of $\sim 3\%$. This then implies that the progenitor of SN~1994I may have been a member of that cluster and, by association, was a massive star.  

\subsection{Pre-supernova mass-loss rate\label{presnMdot}}

The best-fit value for the parameter $K_2$ (see Section \ref{parameterization}) is related to the density of a presupernova stellar wind. If the density distribution in the
circumstellar material (CSM) around SN~1994I is proportional to ${\rm r}^{-2}$, as appropriate for a constant mass-loss rate, constant velocity progenitor wind confirmed by \cite{Immler02}, our measurement of ${\rm K}_2 = 30$ in Table \ref{tbl-4} corresponds to
a CSM external thermal absorption as a function of time of:

\begin{equation}
\label{eq14}
\tau(\lambda, t) = 0.35 ~\lambda^{-2.1} (t/1 ~{\rm day})^{-3}
\end{equation}

Assuming for the moment that the progenitor star for SN~1994I was a red supergiant (RSG), inserting Equation (\ref{eq14}) into equation (16) of Weiler \etal (1986), adopting an
electron temperature of $T_e$ = 20,000\,K for the absorbing CSM gas, and assuming a wind
velocity for this suggested red supergiant progenitor of $w=10$~\kms, we obtain a
mass loss rate of:

\begin{equation}
\label{eq15}
\dot M = 1.8 \times 10^{-7} ~(v_s/20,000 ~{\rm km}~{\rm s}^{-1})^{1.5} ~(w/10 ~{\rm km}~{\rm s}^{-1})~(T_e/20,000~K)^{0.68}~{\rm M}_\odot~{\rm yr}^{-1}
\end{equation}

\noindent This estimate of $\dot M$ is much lower than the estimate of $1.1\times
10^{-5}~(w/10 ~{\rm km}~{\rm s}^{-1})~$~M$_\odot$~yr$^{-1}$ obtained by \cite{Immler02} from their analysis of the SN~1994I X-ray emission data.  

Clumping of the CSM (or a systematic deviation from spherical symmetry) might provide a simple
explanation  for this apparent discrepancy if  most of the ionized gas is
present in the form of large, isolated structures.  In such a case, the mass-loss rate derived from the X-ray emission, which is integrated over the whole CSM,
would be much higher than that estimated from the radio absorption, because the
latter arises only from the material along the line of sight which may not
intercept any dense clumps in the CSM.  Indeed, in deriving their $\dot M$ estimate
\cite{Immler02} assumed that the CSM was both uniform and spherically symmetric.
If such appreciable clumping is present, the average density of the  emitting
material could be  grossly underestimated for radio measurements of absorption along the line-of-sight by a factor equal to the square root
of the clumping factor. The X-ray estimated mass-loss rate (proportional to the ratio of the
X-ray luminosity to the average  density) would then be overestimated by the same
factor.  Conversely, the radio absorption-based estimate would provide only a 
lower limit to the true mass-loss rate. Thus, with a suitable selection of the
clumping factor such
a large difference between the radio and X-ray estimates could be accounted for without abandoning the RSG progenitor assumption. For example, if we take a ``compromise" value of 

\begin{equation}
\label{eq16}
\sqrt{\dot M(radio) \times \dot M(X~ray)}\simeq 1.4 \times 10^{-6} ~(w/10 ~{\rm km}~{\rm s}^{-1})~{\rm M}_\odot~{\rm yr}^{-1}
\end{equation}

\noindent to be appropriate for SN\,1994I progenitor's wind, the clumping factor would be $\sim
40$ and the fraction of mass in the diffuse component $\sim 15$\%.  

On the other hand, if the progenitor of SN~1994I was {\it not} a RSG star but a Wolf-Rayet (WR) star, as has been suggested for Type Ib/c SNe by many authors \citep[see, \eg][ and references therein]{vandenBergh88,Gal-Yam07,Smartt09}, the WR progenitor would have a much higher wind speed ($>10^3 ~{\rm km}~{\rm s}^{-1}$; see, \eg \citealt[][]{Willis96}), as opposed to $\sim 10~{\rm km}~{\rm s}^{-1}$ for a RSG star. Thus, a WR progenitor would revise the above estimate of the presupernova mass-loss rate from Equation \ref{eq16} upwards to $\dot M > 1.4 \times 10^{-4} ~{\rm
M}_\odot~{\rm yr}^{-1}$, a mass-loss rate that is quite possible for a WR star towards the end of its evolution \citep[see, \eg][]{Willis96}.

\section{Conclusions\label{conclusions}}

We present detailed radio observations of SN~1994I at multiple wavelengths from a few days to more than 8 years
after explosion. Although there have been many radio observations of Type Ib/c supernovae since SN~1994I, SN~1994I remains the best studied and most heavily sampled example of the class. Also, even though observations since those of SN~1994I have indicated that synchrotron self-absorption can play a role in the early radio light curves of supernovae that have relatively low density circumstellar matter, such as Type Ib/c and Type IIb, SN~1994I was the first example that was so well sampled early enough that such a conclusion was inescapable.  Also, as seen in other RSNe (see Section \ref{lightcurves}) this low density circumstellar thermal material in SN~1994I may be irregularly distributed as discussed in Section \ref{presnMdot}.

As with previous detailed  observations of the radio emission from supernovae, there is no sign of systematic spectral index evolution with time. However, a changing spectral index with frequency, rather than with time, is a distinct possibility, with the highest observed frequencies perhaps indicating a somewhat flatter spectral index.

As has been noted in the few other radio supernovae which have been sufficiently well studied, the radio light curves, even though their general properties evolve smoothly and are well described by existing models, show significant ``bumps and wiggles.'' SN~1994I shows a significant local ``dip" between $\sim 10$ to $\sim 55$ days after explosion that is interpreted as a temporary decrease in the average density of the circumstellar medium, perhaps implying the origin of SN~1994I in an exploding Wolf-Rayet star in a wide binary system with an early-type stellar companion, although changing mass-loss rate from the progenitor star cannot be ruled out.

The usual scenario for core-collapse (Type Ib/c and Type II) supernovae is that the progenitor was a red supergiant. However, based on the information available from these new observations, it appears that a Wolf-Rayet star undergoing a period of very high mass-loss rate is the most likely progenitor star for SN~1994I. 

\acknowledgements

We are deeply indebted to Dr. Schuyler Van Dyk of the Spitzer Science Center for the large amount of work he did on the data collection and editing, and to Dr.~Barry Clark of the NRAO for scheduling our numerous observations. Many observers also contributed valuable telescope time to enable such a dense sampling of the radio light curves.  KWW wishes to thank the Office of Naval Research (ONR) for the 6.1 funding supporting his research.

\bibliographystyle{apj}

\bibliography{References}

\clearpage

\begin{deluxetable}{cccccccc}
\tabletypesize{\scriptsize}
\tablewidth{6.75in}
\table columns(8}
\tablecaption{Centimeter Flux Density Measurements for SN~1994I \label{tbl-1}}
\tablehead{
\colhead{Obs.} & \colhead{Age} & \colhead{Tel.} & 
\colhead{S(20~cm) $\pm$ rms} & \colhead{S(6~cm) $\pm$ rms} & \colhead{S(3.6~cm) $\pm$ rms} & \colhead{S(2~cm) $\pm$ rms} &  \colhead{S(1.3~cm) $\pm$ rms} \\
\colhead{Date} &  \colhead{(days)} & \colhead{Config.} & \colhead{(mJy)} & \colhead {(mJy)}  & \colhead{(mJy)} & \colhead{(mJy)} & \colhead{(mJy)}
} 
\startdata
31-Mar-94 &= 0.00  & & & & & & \\
03-Apr-94 &  3.167 & VLA-A &                               &                            & 0.554 $\pm$ 0.062  &                              & 2.580 $\pm$ 0.388 \\
04-Apr-94 &  4.120 & VLA-A &                                &                            & 0.737 $\pm$ 0.120  & 2.780  $\pm$ 0.469  & 3.707  $\pm$ 0.891 \\
05-Apr-94 &  5.350 & VLA-A & $<$0.090\tablenotemark{a} & 0.477 $\pm$ 0.091 & 1.743 $\pm$ 0.108  & 6.300 $\pm$ 0.522   & 8.299 $\pm$ 0.891 \\
06-Apr-94 &  6.040 & VLA-A &                               & 0.700 $\pm$ 0.115 &                            &                                &                          \\
06-Apr-94 &  6.135 & Camb   &                               &                             &                            & 10.800 $\pm$ 1.472  &                               \\
06-Apr-94 &  6.483 & VLA-A &                               & 1.400 $\pm$ 0.115 &                            &                               &                             \\
07-Apr-94 &  7.028 & VLA-A &                               &                            &  2.412 $\pm$ 0.152 &                               &                             \\
07-Apr-94 &  7.070 & VLA-A &                               &                            &  3.295 $\pm$ 0.190 &     &                             \\
07-Apr-94 &  7.190 & Camb   &                               &                             &                            & 13.000 $\pm$ 2.385  &                               \\
07-Apr-94 &  7.462 & VLA-A &                               &                            & 3.656 $\pm$ 0.194  & 13.100 $\pm$ 1.062  & 16.300 $\pm$ 1.676   \\
08-Apr-94 &  8.020 & Camb   &                               &                             &                             & 14.500  $\pm$ 2.086 &                             \\
08-Apr-94 &  8.288 & VLA-A & $<$0.207\tablenotemark{a} & 1.250 $\pm$ 0.153  & 4.990 $\pm$ 0.250 & 14.698 $\pm$ 1.202   & 15.458 $\pm$ 1.673 \\
09-Apr-94 &  9.180 & Camb   &                               &                             &                            & 16.000  $\pm$ 2.561 &                                \\
09-Apr-94 &  9.386 & VLA-A &                              & 2.020 $\pm$ 0.144  & 6.514 $\pm$ 0.334 & 17.200 $\pm$ 1.409  & 17.100  $\pm$ 1.810  \\
10-Apr-94 & 10.561 & VLA-A & $<$0.225\tablenotemark{a} & 2.123 $\pm$ 0.176  & 8.930 $\pm$ 0.455 & 19.171 $\pm$ 1.488  & 18.593  $\pm$ 2.028  \\
11-Apr-94 & 11.180 & Camb   &                             &                             &                             & 17.500  $\pm$ 2.658 &                                 \\
12-Apr-94 & 12.180 & Camb   &                             &                              &                             & 17.200  $\pm$ 2.282  &                                 \\
12-Apr-94 & 12.395 & VLA-A &  $<$0.132\tablenotemark{a} &  4.120 $\pm$ 0.221 & 11.43 $\pm$ 0.577 &   15.860 $\pm$ 1.249   & 12.400 $\pm$ 1.349   \\
13-Apr-94 & 13.220 & Camb   &                             &                             &                               & 13.900  $\pm$ 2.436 &                               \\
13-Apr-94 & 13.372 & VLA-A & $<$0.162\tablenotemark{a} & 4.643 $\pm$ 0.271   &12.605 $\pm$ 0.646  & 17.178  $\pm$ 1.365  & 12.610  $\pm$ 1.414 \\
14-Apr-94 & 14.220 & Camb   &                             &                             &                              & 15.600  $\pm$ 2.536 &                                  \\
14-Apr-94 & 14.367 & VLA-A & 0.300 $\pm$ 0.067  & 5.053 $\pm$ 0.267  & 13.000 $\pm$ 0.660 &  16.500 $\pm$ 1.382 &   \\
15-Apr-94 & 15.220 & Camb   &                             &                             &                              & 15.000  $\pm$ 2.500 &                                   \\
15-Apr-94 & 15.360 & VLA-A & $<$0.258\tablenotemark{a} & 5.587 $\pm$ 0.292& 14.500 $\pm$ 0.733  & 16.500 $\pm$ 1.364 & 10.500  $\pm$ 1.178     \\
16-Apr-94 & 16.220 & Camb  &                             &                               &                              & 16.500  $\pm$ 2.593 &                                    \\
16-Apr-94 & 16.356 & VLA-A & 0.370  $\pm$ 0.095 & 6.451  $\pm$ 0.338 &16.200 $\pm$ 0.820 & 15.600  $\pm$ 1.236 & 11.000  $\pm$ 1.252 \\
17-Apr-94 & 17.210 & Camb    &                            &                             &                              & 16.100  $\pm$ 2.200 &                              \\
17-Apr-94 & 17.363 & VLA-A & 0.463  $\pm$ 0.105 &  7.664  $\pm$ 0.400 & 17.200 $\pm$ 0.871 & 16.900  $\pm$ 1.328 & 10.600  $\pm$ 1.185  \\
18-Apr-94 & 18.199 & VLA-A &                             &  7.390  $\pm$ 0.392 & 15.770 $\pm$ 0.810 & 15.743  $\pm$ 1.215 & 10.590  $\pm$ 1.205  \\
18-Apr-94 & 18.210 & Camb    &                             &                            &                               & 15.900  $\pm$ 2.555 &                               \\
19-Apr-94 & 19.220 & Camb    &                            &                             &                               & 14.000  $\pm$ 2.441 &                                \\
20-Apr-94 & 20.200 & Camb    &                            &                             &                               & 13.500  $\pm$ 2.018 &                                \\
20-Apr-94 & 20.494 & VLA-A & 0.421  $\pm$ 0.107 & 9.250  $\pm$ 0.493 &16.483  $\pm$ 0.850  & 8.757  $\pm$ 0.960  & 5.673  $\pm$ 1.185     \\
21-Apr-94 & 21.200 & Camb    &                            &                              &                              & 14.000  $\pm$ 2.052 &                                 \\
22-Apr-94 & 22.210 & Camb    &                           &                               &                               & 13.000  $\pm$ 1.985 &                                \\
23-Apr-94 & 23.056 & VLA-A & 0.386  $\pm$ 0.089 &                              &                               &                              &                                  \\
23-Apr-94 & 23.150 & Camb   &                            &                               &                                & 12.000  $\pm$ 1.921 &                                 \\
24-Apr-94 & 24.336 & VLA-A & 0.688  $\pm$ 0.125 & 11.130  $\pm$ 0.568 &                            &                             &                                    \\
25-Apr-94 & 25.140 & Camb    &                           &                                &                            & 10.000  $\pm$ 1.414 &                                      \\
25-Apr-94 & 25.333 & VLA-A & 0.720 $\pm$ 0.122 & 10.760  $\pm$ 0.551  & 12.300  $\pm$ 0.651 &   & 4.620  $\pm$ 0.816  \\
28-Apr-94 & 28.163 & VLA-A & 0.683  $\pm$ 0.115 & 13.713  $\pm$ 0.693 &14.723  $\pm$ 0.760 & 7.503  $\pm$ 0.733 & 4.990  $\pm$ 0.706    \\
29-Apr-94 & 29.200 & Camb    &                            &                               &                               & 10.000  $\pm$ 1.281 &                            \\
30-Apr-94 & 30.320 & VLA-A & 1.618  $\pm$ 0.167 & 14.063  $\pm$ 0.726 &14.650  $\pm$ 0.757 & 8.720  $\pm$ 0.848 & 6.293  $\pm$ 0.971 \\
02-May-94 & 32.140 & Camb  &                             &                               &                              &  9.800  $\pm$ 1.400 &                             \\
03-May-94 & 33.130 & Camb  &                             &                               &                              &  9.100  $\pm$ 1.212 &                            \\
03-May-94 & 33.279 & VLA-BnA & 1.170  $\pm$ 0.142 & 16.823  $\pm$ 0.847 &11.440  $\pm$ 0.591 &  & 3.988  $\pm$ 0.624 \\
10-May-94 & 40.228 & VLA-BnA & 2.660  $\pm$ 0.267 & 17.550  $\pm$ 0.880 &13.040  $\pm$ 0.656 & 5.330  $\pm$ 0.477 & 2.180  $\pm$ 0.555 \\
12-May-94 & 42.090 & Camb  &                             &                               &                              &  8.000  $\pm$ 1.281 &                            \\
15-May-94 & 45.179 & VLA-BnA & 2.470  $\pm$ 0.270 & 16.995  $\pm$ 0.861 &12.788  $\pm$ 0.652 & 6.555  $\pm$ 0.615 & 3.533  $\pm$ 0.662 \\
16-May-94 & 46.100 & Camb  &                             &                               &                              &  8.000  $\pm$ 1.281 &                            \\
17-May-94 & 47.100 & Camb  &                             &                               &                              &  8.400  $\pm$ 1.306 &                            \\
19-May-94 & 49.252 & VLA-BnA & 3.420 $\pm$ 0.343  & 15.940 $\pm$ 0.798  &11.040 $\pm$ 0.554  & 5.610 $\pm$ 0.523   & 2.810 $\pm$ 0.539 \\
23-May-94 & 53.208 & VLA-BnA &                             & 15.860  $\pm$ 0.809 &11.305  $\pm$ 0.570 & 5.380  $\pm$ 0.547  & 2.680  $\pm$ 0.722 \\
24-May-94 & 54.060 & Camb  &                             &                               &                              &  5.500  $\pm$ 1.141 &                            \\
27-May-94 & 57.060 & Camb  &                             &                               &                              &  5.500  $\pm$ 0.890 &                            \\
28-May-94 & 58.136 & VLA-BnA & 4.660  $\pm$ 0.466 & 14.030 $\pm$ 0.704 & 8.400 $\pm$ 0.425 & 4.240  $\pm$ 0.473 & 2.820  $\pm$ 0.645  \\
01-Jun-94 & 62.188 & VLA-BnA & 5.080 $\pm$ 0.508 & 12.400 $\pm$ 0.626 & 7.670  $\pm$ 0.392 & 3.390 $\pm$ 0.441 &  $<$2.250\tablenotemark{a}  \\
17-Jun-94 & 78.165 & VLA-B & 7.190  $\pm$ 0.729 & 9.640  $\pm$ 0.497 & 5.860  $\pm$ 0.310 & 2.920  $\pm$ 0.404 & 1.910  $\pm$ 0.356 \\
29-Jun-94 & 90.173 & VLA-B & 10.940 $\pm$ 1.103 & 7.620 $\pm$ 0.388 & 4.240 $\pm$ 0.221 & 2.420 $\pm$ 0.359 & 2.050 $\pm$ 0.596 \\
07-Jul-94 & 98.027 & VLA-B & 10.530 $\pm$ 1.065 &7.340 $\pm$ 0.369 & 4.000 $\pm$ 0.207  & 2.150 $\pm$ 0.323 & $<$1.620\tablenotemark{a} \\
08-Aug-94 &130.937 & VLA-B &9.410 $\pm$ 0.951& 5.020 $\pm$ 0.262&2.750 $\pm$ 0.150&1.480 $\pm$ 0.348 & $<$2.400\tablenotemark{a} \\
22-Aug-94 &144.959 & VLA-B &11.480  $\pm$ 1.155&3.760  $\pm$ 0.234&1.930  $\pm$ 0.154&1.620  $\pm$ 0.476&$<$1.980\tablenotemark{a} \\
01-Sep-94 &154.753 & VLA-B &9.250  $\pm$ 1.277&2.073  $\pm$ 0.174&1.415  $\pm$ 0.139&$<$1.320\tablenotemark{a}& $<$4.050\tablenotemark{a} \\
30-Sep-94&183.774&VLA-CnB&10.510$\pm$1.196&2.824$\pm$0.213&1.530  $\pm$ 0.097&$<$1.830\tablenotemark{a}&$<$2.160\tablenotemark{a} \\
13-Oct-94&196.687&VLA-C&$<$2.841\tablenotemark{a}&2.203$\pm$0.264&1.117$\pm$0.169&$<$1.140\tablenotemark{a}&$<$1.917\tablenotemark{a} \\
07-Nov-94 &221.513 & VLA-C &  & 1.257  $\pm$ 0.191 & 1.220  $\pm$ 0.117 & $<$0.990\tablenotemark{a}& $<$2.970\tablenotemark{a} \\
05-Dec-94 &249.708 & VLA-D &   &   2.209  $\pm$ 0.357   & 1.024  $\pm$ 0.176  & $<$2.220\tablenotemark{a}     &  \\
05-Jan-95 &280.507 & VLA-D &  &    &  & $<$1.650\tablenotemark{a}     &    \\
06-Apr-95 &371.173 & VLA-D &   &    &  & $<$1.032\tablenotemark{a}&  $<$1.920\tablenotemark{a}  \\
16-Jun-95 &442.147 & VLA-DnA & 2.397 $\pm$ 0.253 & 0.855  $\pm$ 0.102   & &  &  \\
06-Oct-95 &554.790&VLA-BnA&1.087  $\pm$ 0.121&0.503  $\pm$ 0.113&0.391  $\pm$ 0.072&$<$0.636\tablenotemark{a}  &  \\
12-Jan-96 &652.500 & VLA-CnB & & 0.460  $\pm$ 0.046 &  & $<$0.480\tablenotemark{a}  &  \\
05-Oct-96 &919.847 & VLA-DnA & 0.580 $\pm$ 0.152 & 0.250  $\pm$ 0.071 &  $<$0.390\tablenotemark{a} &  &   \\
13-Jun-99 & 1900.000   & VLA-DnA & $<$0.222\tablenotemark{a}  &     &   &   \\
05-Apr-02 & 2927.000  & VLA-A     & 0.150 $\pm$ 0.026 &    &  &  &   \\
\enddata
\tablenotetext{a}{All upper limits are 3$\sigma$}
\end{deluxetable}

\clearpage

\begin{deluxetable}{cccccc}
\tablecaption{Millimeter Flux Density Measurements for SN~1994I \label{tbl-2}}
\tabletypesize{\scriptsize}
\tablewidth{6.0in}
\table columns(6}
\tablehead{
\colhead{Obs.} & \colhead{Age} & \colhead{Tel.} &  \colhead{Flux Density (S) $\pm$ rms\tablenotemark{a}} & \colhead{Frequency} & \colhead{Reference}  \\
\colhead{Date} &  \colhead{(days)} &  &  \colhead{(mJy)} & \colhead {(GHz)} & \colhead {}  } 
\startdata
31-Mar-94 &= 0.00  & & &  & \\
05-Apr-94 &  5.45 & OVRO & 11.0 $\pm$ 0.6                &    99      & \cite{PhillipsJ94}   \\
06-Apr-94 &  6.45 & OVRO &    $<$12\tablenotemark{b}              & 218  & \cite{PhillipsJ94}          \\
\enddata
\tablenotetext{a}{Quoted errors are taken from the original references.}
\tablenotetext{b}{All upper limits are 3$\sigma$}
\end{deluxetable}

\clearpage

\begin{deluxetable}{cccccc}
\tabletypesize{\scriptsize}
\tablewidth{4.0in}
\table columns(6}
\tablecaption{VLA Calibrator Flux Density Measurements\tablenotemark{a} \label{tbl-3}}
\tablehead{
\colhead{Obs.} & \colhead{S(20~cm)} & \colhead{S(6~cm)} & \colhead{S(3.6~cm)} & \colhead{S(2~cm)} &  \colhead{S(1.3~cm)} \\
\colhead{Date} &  \colhead{(mJy)} & \colhead {(mJy)}  & \colhead{(mJy)} & \colhead{(mJy)} & \colhead{(mJy)}
} 
\startdata
03-Apr-94 &                                &                            & 0.539\tablenotemark{b} &   & 0.338\tablenotemark{b} \\
04-Apr-94 &                                &                            & 1.014\tablenotemark{c} & 0.953\tablenotemark{c}  & 0.830\tablenotemark{c} \\
05-Apr-94 & 0.904 & 1.038 & 1.014  & 0.953   & 0.830 \\
06-Apr-94 &                               & 1.020\tablenotemark{c} &                            &                               &                             \\
07-Apr-94 &                               &                            & 0.994 & 0.959\tablenotemark{c}  & 0.880\tablenotemark{c}   \\
08-Apr-94 & 0.899\tablenotemark{c} & 1.025\tablenotemark{c}  & 1.000\tablenotemark{c} & 0.954\tablenotemark{c} & 0.875\tablenotemark{c}  \\
09-Apr-94 &                              & 1.040\tablenotemark{c}  & 1.019\tablenotemark{c}  & 0.980\tablenotemark{c}  & 0.910\tablenotemark{c}  \\
10-Apr-94 & 0.894 & 1.013  & 0.987 & 0.956  & 0.920  \\
12-Apr-94 &  0.887\tablenotemark{c} &  1.006\tablenotemark{c} & 0.974\tablenotemark{c} &  0.940\tablenotemark{c}  & 0.864\tablenotemark{c}  \\
13-Apr-94 & 0.884 & 1.017 &0.993  & 0.955  & 0.865 \\
14-Apr-94 & 0.875\tablenotemark{c}  & 1.026  & 1.027\tablenotemark{c} &  0.976\tablenotemark{c} &   \\
15-Apr-94 & 0.875\tablenotemark{c} & 1.017\tablenotemark{c} & 1.029\tablenotemark{c}   & 0.993\tablenotemark{c} & 0.839\tablenotemark{c}  \\
16-Apr-94 & 0.875\tablenotemark{c} & 1.017\tablenotemark{c} & 1.029\tablenotemark{c} & 0.993\tablenotemark{c} & 0.839\tablenotemark{c} \\
17-Apr-94 & 0.875\tablenotemark{c} &  1.060\tablenotemark{c} & 1.032\tablenotemark{c} & 0.976\tablenotemark{c} & 0.810\tablenotemark{c}  \\
18-Apr-94 &                     &  0.984      &  0.963 & 0.930& 0.901  \\
20-Apr-94 & 0.875\tablenotemark{c} & 1.005\tablenotemark{c} & 0.964\tablenotemark{c} & 0.918\tablenotemark{c} & 0.773   \\
23-Apr-94 & 0.875\tablenotemark{c} &                              &                               &                              &                                  \\
24-Apr-94 & 0.875\tablenotemark{c} & 1.015\tablenotemark{c} &                            &                             &                                    \\
25-Apr-94 & 0.904 & 1.015  & 0.946 &   & 0.689  \\
28-Apr-94 & 0.866 & 0.995 & 0.937 & 0.881 & 0.683    \\
30-Apr-94 & 0.887\tablenotemark{c} & 0.993\tablenotemark{c} & 0.958\tablenotemark{c} & 0.890\tablenotemark{c} & 0.792\tablenotemark{c} \\
03-May-94 & 0.905 & 0.992 & 0.940 &  & 0.900 \\
10-May-94 & 0.890 & 0.946 & 0.930 & 0.859 & 0.846 \\
15-May-94 & 0.899 & 0.983 & 0.949 & 0.900 & 0.816 \\
19-May-94 & 0.868  & 0.948  & 0.920  & 0.830   & 0.764 \\
23-May-94 &                             & 0.986 & 0.933 & 0.863  & 0.797 \\
28-May-94 & 0.872 & 0.976 & 0.939 & 0.878 & 0.822  \\
01-Jun-94 & 0.894 & 0.996 & 0.963 & 0.874 &  0.905  \\
17-Jun-94 & 0.894 & 0.991 & 0.957 & 0.875 & 0.793 \\
29-Jun-94 & 0.883 & 0.958 & 0.918 & 0.829 & 0.786 \\
07-Jul-94 & 0.882 & 0.927 & 0.885 & 0.783 & 0.753  \\
08-Aug-94 & 0.881 & 0.912 & 0.879 & 0.846  & 0.827  \\
22-Aug-94  & 0.926 & 0.863 & 0.784 & 0.635 & 0.633  \\
01-Sep-94 & 0.906\tablenotemark{c} & 0.852\tablenotemark{c} & 0.807\tablenotemark{c} & 0.657 &   0.665 \\
30-Sep-94 & 0.885 & 0.831 & 0.830 & 0.654 & 0.528  \\
13-Oct-94 & 0.840\tablenotemark{c} & 0.834\tablenotemark{c} & 0.822\tablenotemark{c} & 0.705\tablenotemark{c} & 0.610\tablenotemark{c}   \\
07-Nov-94 &                                 & 0.826 & 0.812 & 0.755 & 0.693  \\
05-Dec-94  &                               &   0.832 & 0.782 & 0.680 &                \\
05-Jan-95 &    &    &   & 0.726 &    \\
06-Apr-95 &   &  &    & 0.697  &  0.634 \\
16-Jun-95 & 0.760 & 0.764  &  &   &  \\
06-Oct-95 & 0.705 &  0.750 & 0.738 & 0.702 &   \\
12-Jan-96 &    & 0.726 &    & 0.697 &    \\
05-Oct-96 & 0.622 & 0.775 &  0.792  &    &  \\
13-Jun-99 & 0.498  &    &   &  &  \\
05-Apr-02  & 0.537\tablenotemark{d} &   &  & &  \\
\enddata
\tablenotetext{a}{Unless otherwise noted, the flux density measurements are for J1418+546.}
\tablenotetext{b}{Flux density measurements are for J1415+463}
\tablenotetext{c}{Interpolated from nearby measurements}
\tablenotetext{d}{Flux density measurement is for 1327+434}
\end{deluxetable}

\clearpage

\begin{deluxetable}{ccc}
\tabletypesize{\footnotesize}
\tablecaption{Best fit parameters for SN~1994I and Comparison with SN~1993J. \label{tbl-4}}
\tablewidth{0pt}
\tablehead{
\colhead{Parameter} & \colhead{SN~1994I} & \colhead{SN~1993J\tablenotemark{a}}}
\startdata
$K_1$                   & $4.78 \times 10^3$       & $4.8\times 10^3$          \\
$\alpha$                & $-1.22$                  & $-0.81$                               \\
$\beta$                 & $-1.42$                 & $-0.73$                    \\ 
$K_2$                   & $3.0 \times 10^{1}$                & $1.6 \times 10^2$                \\
$\delta$                & $$\tbond$$$-3.00$\tablenotemark{b}                       &   $-1.88$                       \\
$K_3$                   & $\tbond$0\tablenotemark{b}                & $4.3 \times 10^5$           \\
$\delta^{\prime}$       & --                       & $-2.83$                         \\ 
$K_4$                   & $4.55 \times 10^{-2}$    & $\tbond$0       \\
$K_5$                   & $6.90 \times 10^5$       & $2.62 \times 10^3$                            \\
${\delta}^{\prime\prime}$ & $-4.08$                & $-2.05$                           \\
$K_6$                   & $9.96 \times 10^3$       & $\tbond$0                         \\
${\delta}^{\prime\prime\prime}$ & $-2.63$          & --    \\ 
$\rm {Distance}$ \ (Mpc) & 8.9      & 3.63          \\
$\rm {Date~of~explosion}$ \ (t$_0$) & 1994~March~31      & 1993~March~28          \\
$\rm {Time~ to} ~S_{\rm 6\ cm\ peak}$ \ (days) & 34.4      & 133          \\
$S_{\rm 6\ cm\ peak}$\ (mJy) & $16.2$  &  $96.9$         \\
$L_{\rm 6\ cm\ peak}$\ ($\rm erg ~\rm s^{-1} ~\rm Hz^{-1}$) & $1.5\times 10^{27}$  &  $1.5\times 10^{27}$         \\
$\dot M$ (${\rm M}_{\odot} \rm~ {\rm yr}^{-1}$) & $>1.4 \times 10^{-4}$\tablenotemark{c}  &  $0.5-5.9\times 10^{-6}$\tablenotemark{d}          \\

\enddata

\tablenotetext{a}{Weiler \etal 2007}
\tablenotetext{b}{Adopted. Because the amount of thermal material surrounding SN~1994I is so low, it is not possible to determine $\delta$ and $K_3$  from the data fitting and nominal values were adopted.}
\tablenotetext{c}{Under the assumption of a Wolf-Rayet progenitor; see Section \ref{presnMdot}}.
\tablenotetext{d}{Weiler \etal (2007) found that the mass-loss rate from SN~1993J increased from $\sim 5.4\times 10^{-7}~M_{\odot} \rm~ yr^{-1}$ at the time of shock breakout to  $\sim 5.9\times 10^{-6}~M_{\odot} \rm~ yr^{-1}$around day $\sim 3100$ ($\sim 8000$~yr before explosion).}
\end{deluxetable}

\clearpage

\figcaption[]{The radio light curves for SN~1994I for: top left 0.3 cm, top right 1.2 cm, middle left 2 cm, middle right 3.6 cm, lower left 6 cm, lower right 20 cm. The lines on all figures represent the best fit model as described in the text with the parameters listed in Table 4.  \label{fig-1}}

\figcaption[]{The spectral index evolution for SN~1994I for: top left 1.2-2 cm, top right 2-3.6 cm, lower left 3.6-6 cm, lower right 6-20 cm. The lines on all figures represent the best fit model as described in the text with the parameters listed in Table 4.  \label{fig-2}}

\figcaption[]{(left) The brightness temperature (T$_B$) evolution, uncorrected for thermal absorption, for SN~1994I for, top left 1.2 cm, top right 2 cm, middle left 3.6 cm, middle right 6 cm, and lower left 20 cm and (right)  the brightness temperature (T$_B$) evolution, uncorrected for thermal absorption, for SN~1993J \citep{Weiler07} for, from left to right, 0.3 cm (cross), 1.2 cm (filled square), 2 cm (open square), 3.6 cm (filled circle), 6 cm (open triangle), 20 cm (filled triangle), 49 cm (star), and 90 cm (open diamond).  The horizontal dashed lines in all plots denote the limiting value of T$_B\simeq 3\times 10^{11}$~K \citep{Kellermann69,Readhead94}.\label{fig-3}}

\figcaption[]{Cartoon (taken from \cite{Weiler02}), not to scale, of the supernova and its shocks, along with the stellar wind established circumstellar medium (CSM), the interstellar medium (ISM), and more distant ionized hydrogen (HII) absorbing gas. The radio emission is thought to arise near the blastwave front.  The expected locations of the several absorbing terms in Equations (\ref{eq5} -- \ref{eq12}) are illustrated. \label{fig-4}}

\figcaption[]{(left) Log of the deviations of the ratio of observed to model flux density for SN~1994I for the best fit model listed in Table \ref{tbl-4} from 1.2 cm (top) to 20 cm (bottom) and (right) composite of the deviations at all wavelengths shown on the left, with a smoothed curve added to guide the eye.\label{fig-5}}

\clearpage

\begin{figure}
\figurenum{fig-1}
\epsscale{1.0}
\plotfiddle{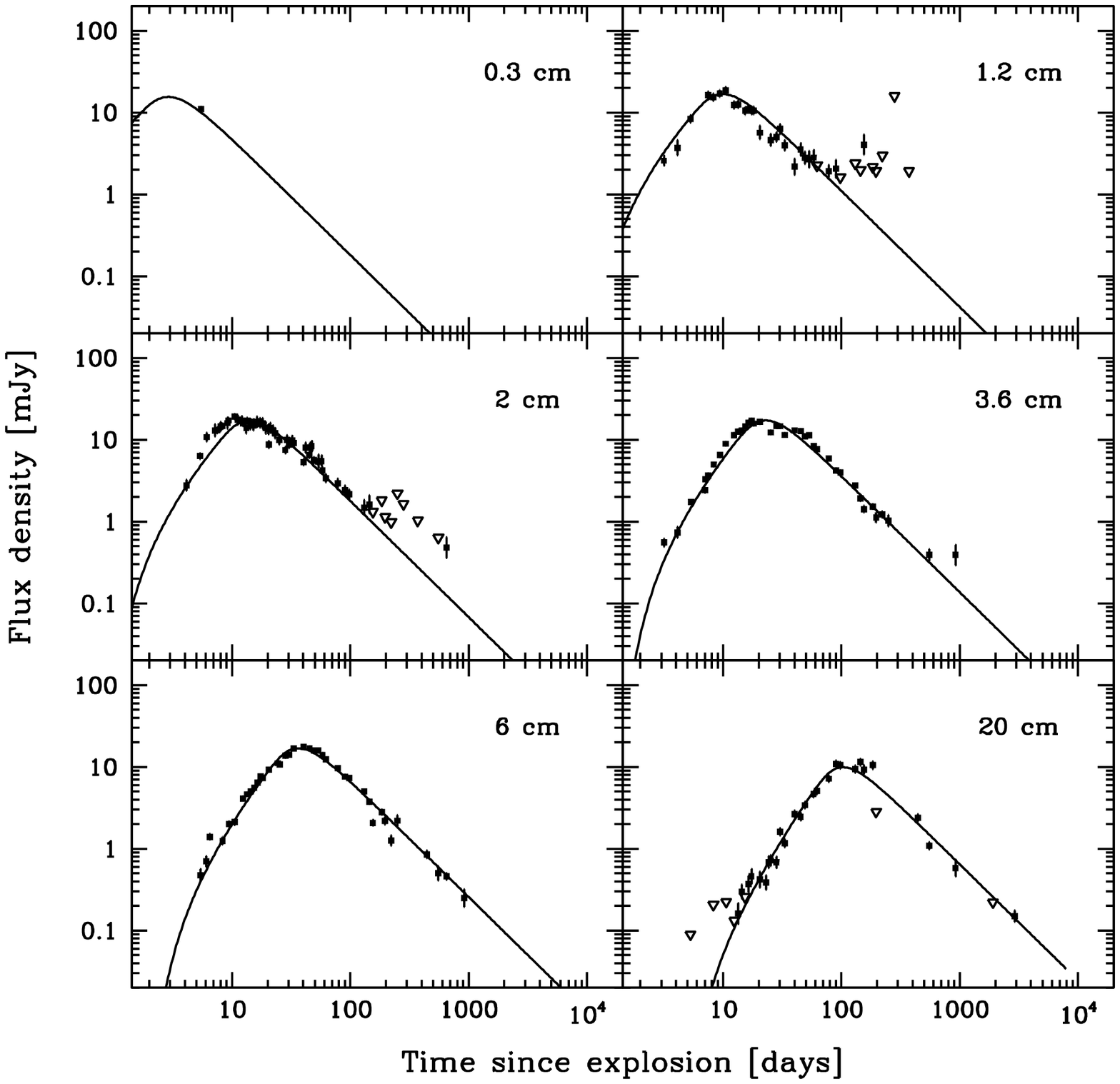}{0.0in}{90.}{500.}{500.}{0}{0}
\end{figure} 

\clearpage

\begin{figure}
\figurenum{fig-2}
\epsscale{1.0}
\plotfiddle{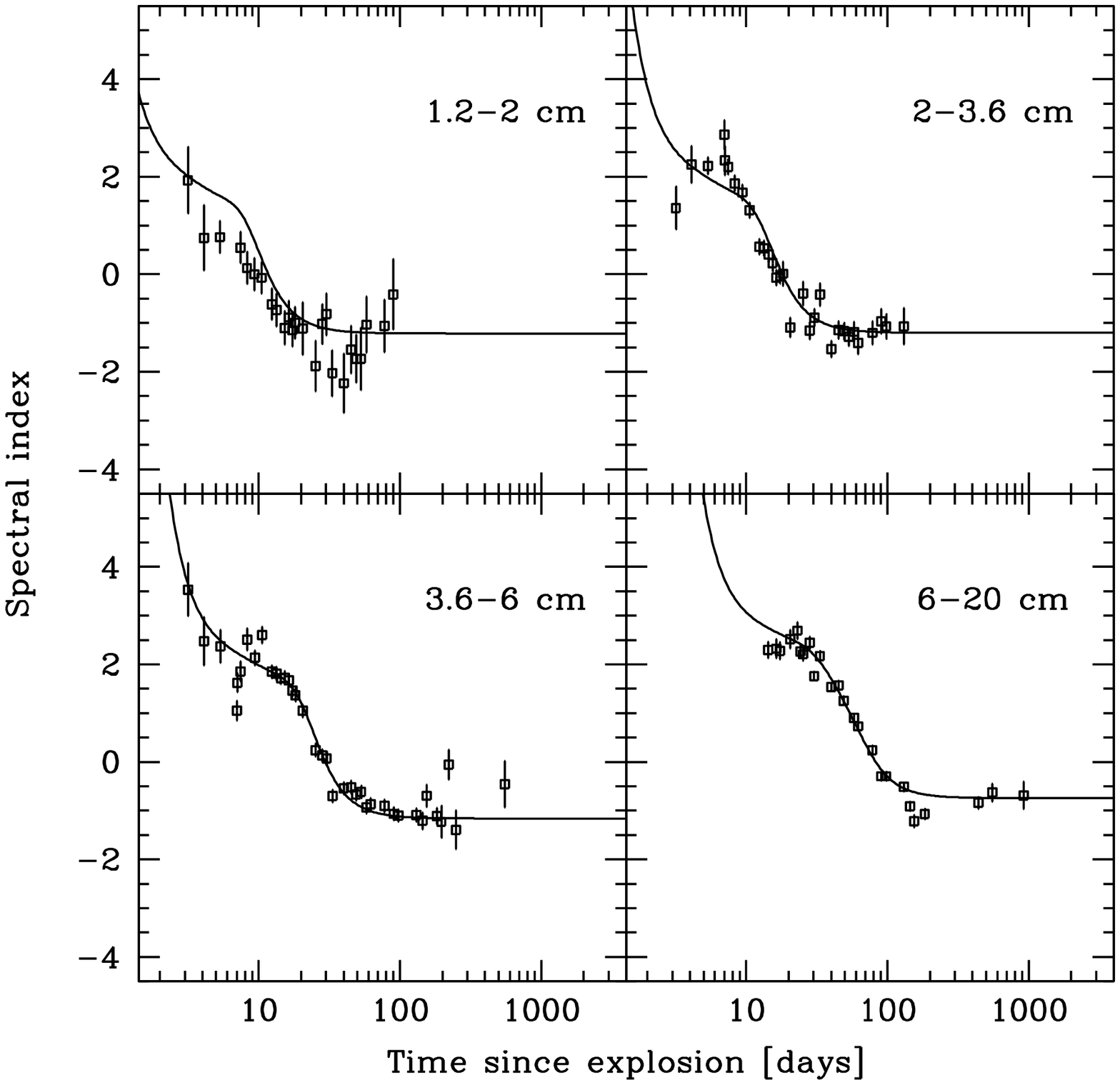}{0.0in}{90.}{500.}{500.}{0}{0}
\end{figure} 

\clearpage

\begin{figure}
\figurenum{fig-3}
\epsscale{1.0}
\plotfiddle{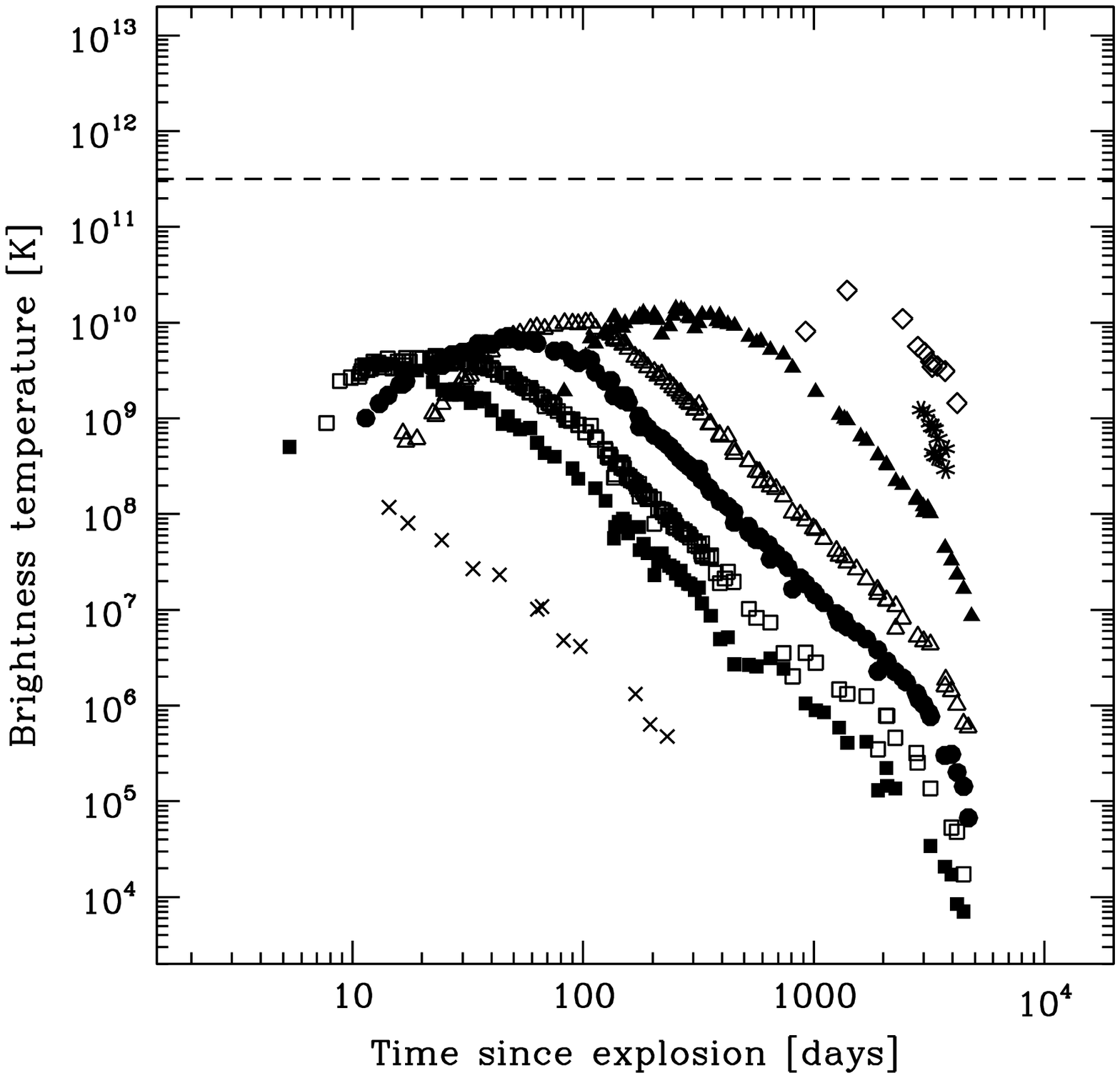}{0.0in}{90.}{325.}{325.}{-20}{0}
\plotfiddle{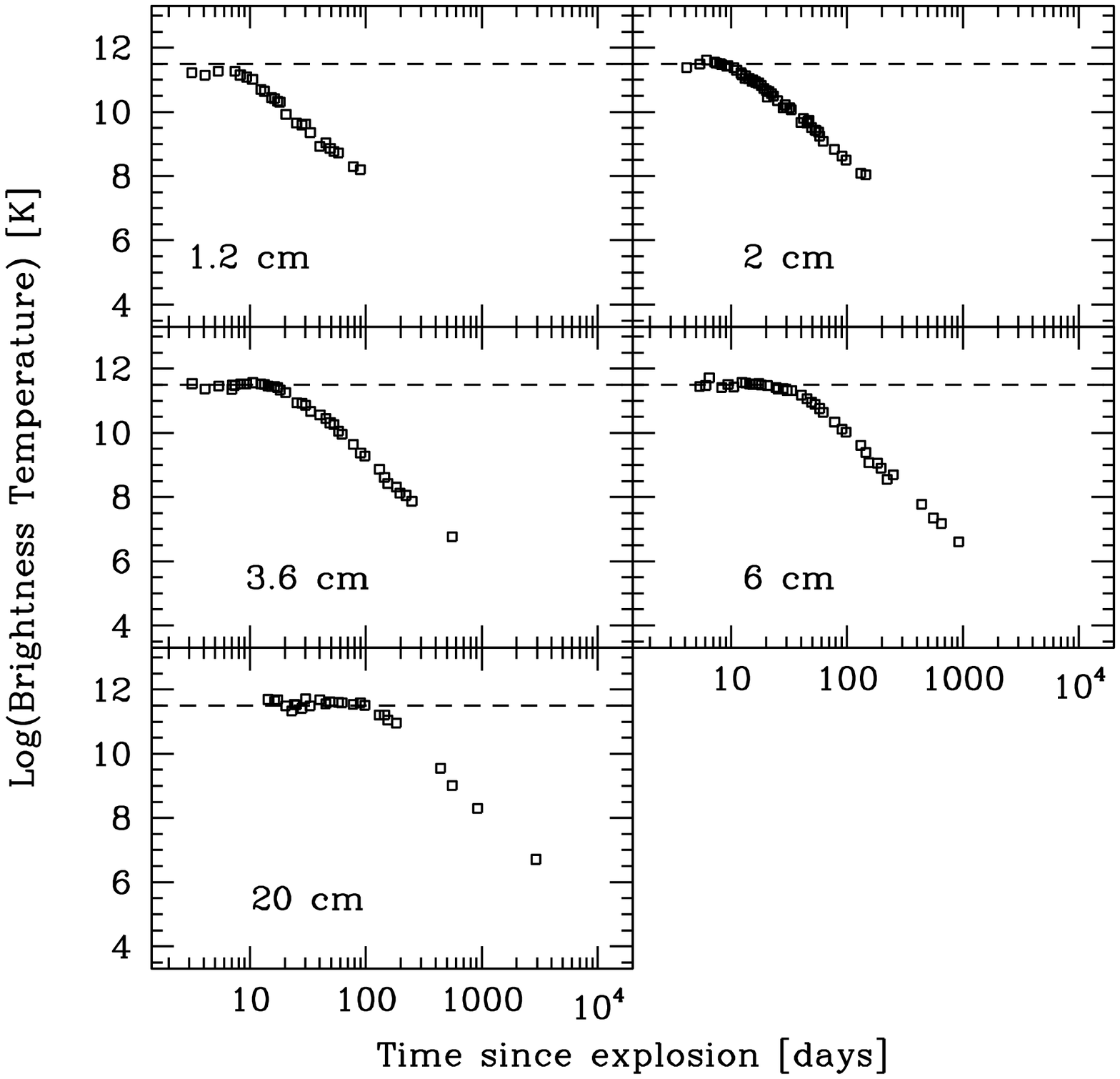}{0.0in}{90.}{325.}{325.}{0}{0}
\end{figure} 

\clearpage

\begin{figure}
\figurenum{fig-4}
\epsscale{1.0}
\plotfiddle{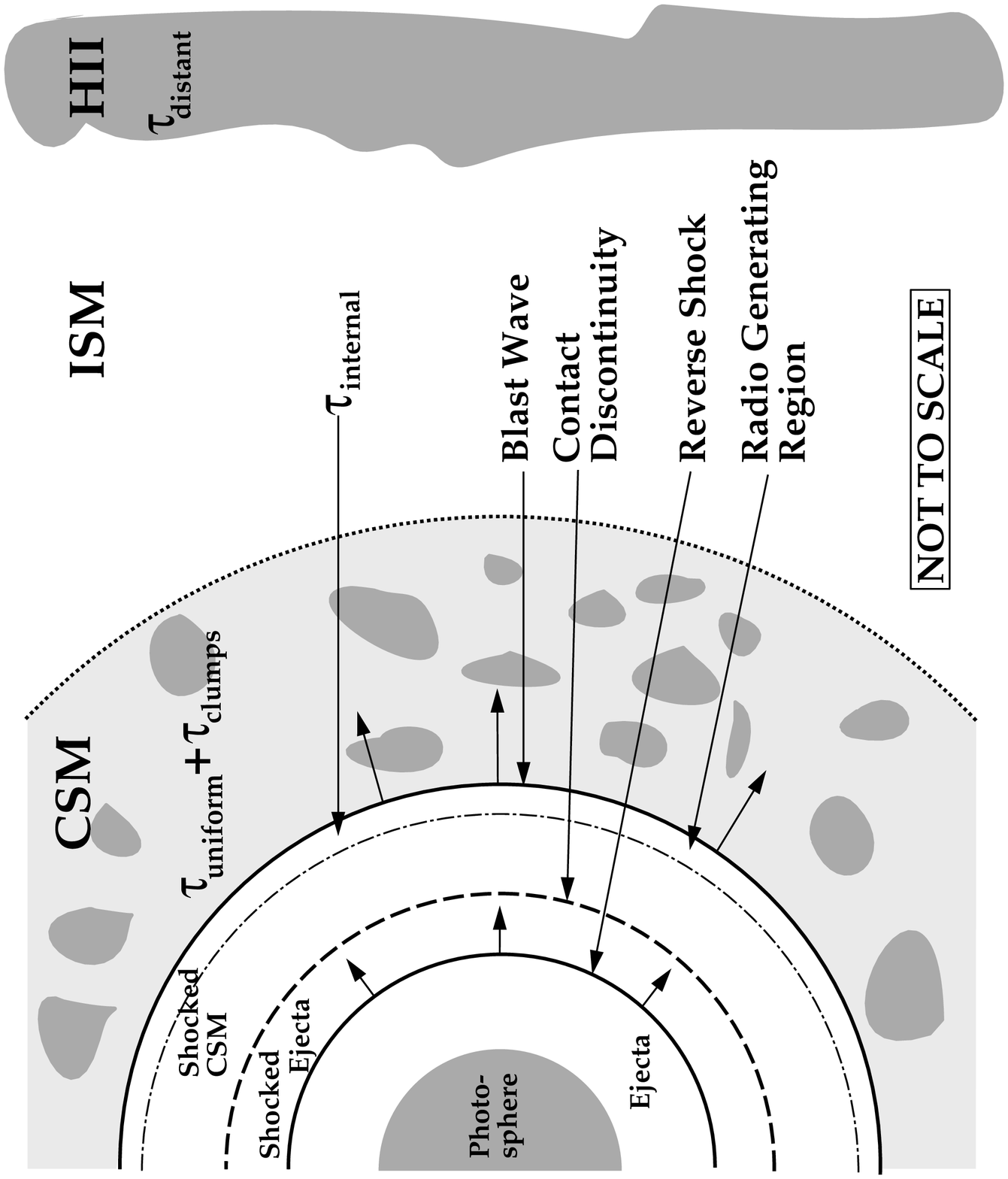}{0.0in}{0.}{500.}{500.}{0}{0}
\end{figure} 

\clearpage

\begin{figure}
\figurenum{fig-5}
\epsscale{1.0}
\plotfiddle{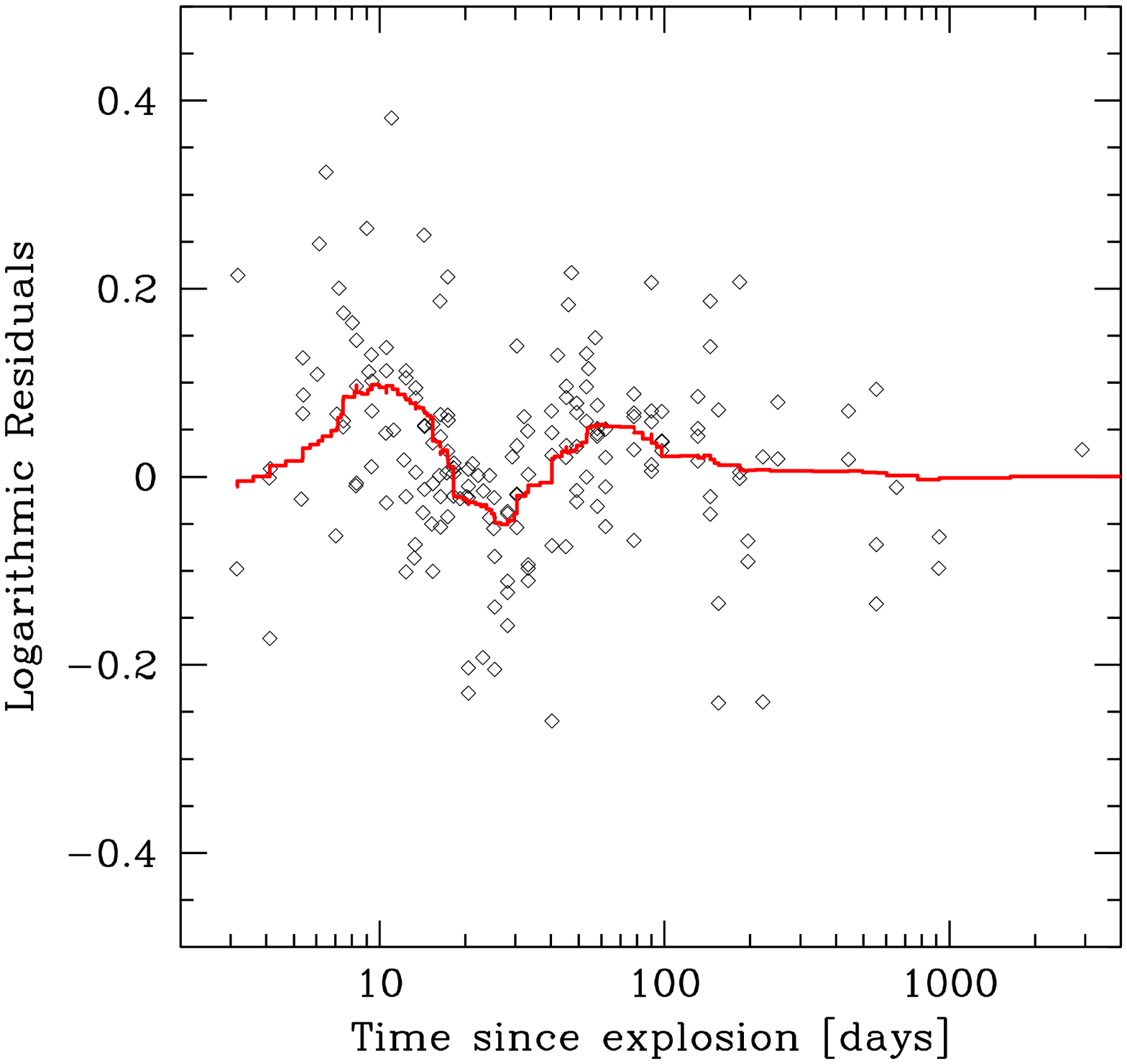}{0.0in}{90.}{330.}{330.}{0}{0}
\plotfiddle{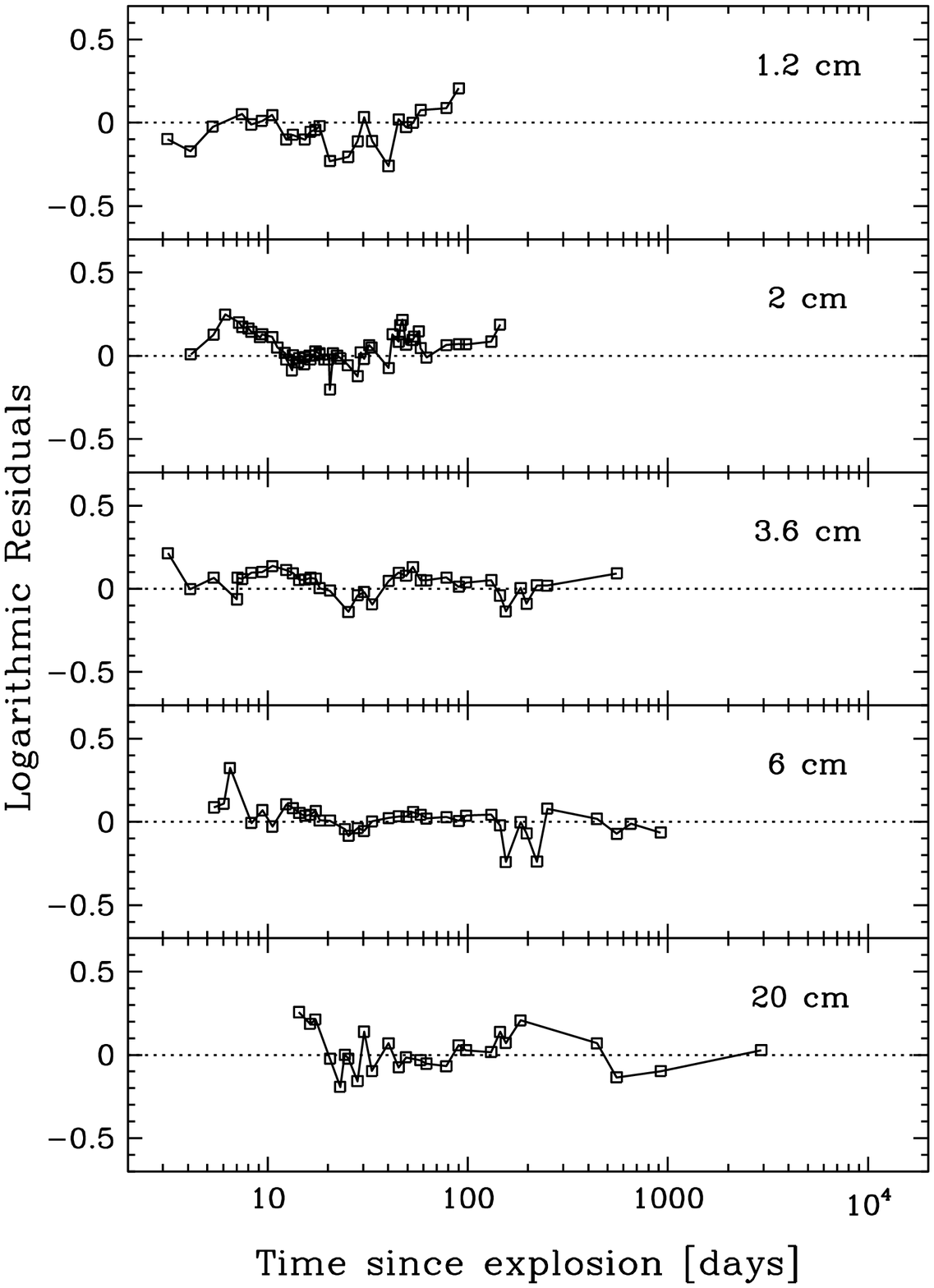}{0.0in}{90.}{330.}{330.}{00}{0}
\end{figure} 

\end{document}